# Review of Data Integrity Attacks and Mitigation Methods in Edge computing

Poornima Mahadevappa[1][0000-0001-9414-3464] and Raja Kumar Murugesan[1][0000-0001-9500-1361]

[1] Taylor's University, Subang Jaya, Malaysia
poornimamahadevappa@sd.taylors.edu.my
rajakumar.murugesan@taylors.edu.my

**Abstract.** In recent years, edge computing has emerged as a promising technology due to its unique feature of real-time computing and parallel processing. They provide computing and storage capability closer to the data source and bypass the distant links to the cloud. The edge data analytics process the ubiquitous data on the edge layer to offer real-time interactions for the application. However, this process can be prone to security threats like gaining malicious access or manipulate sensitive data. This can lead to the intruder's control, alter, or add erroneous data affecting the integrity and data analysis efficiency. Due to the lack of transparency of stakeholders processing edge data, it is challenging to identify the vulnerabilities. Many reviews are available on data security issues on the edge layer; however, they do not address integrity issues exclusively. Therefore, this paper concentrates only on data integrity threats that directly influence edge data analysis. Further shortcomings in existing work are identified with few research directions.

## 1 Introduction

Edge computing is the advancement of cloud computing by bringing computation and networking ability closer to edge devices. This extended computing paradigm has addressed the challenges of increased digitalization due to billions of ubiquitous devices. Edge computing facilitates the resources to ubiquitous devices according to user requirements in real-time and reduces the burden of voluminous data on the cloud. This is achieved by the highly virtualized edge layer between the cloud and ubiquitous devices providing network, storage and compute services [1]. Currently, various applications in healthcare, VANET (Vehicular Adhoc Network), smart grid, smart transport, smart cities, augmented reality are adopting edge computing and the inability of cloud to meet stringent latency requirements is achieved [2–5]. It is forecasted that by 2025 edge computing market revenue will be up to 28 billion U.S dollars[1]. As evidence, localized computing power on edge can grow exponentially soon.

---

[1] https://www.statista.com/statistics/948744/worldwide-edge-computing-market-size/

Edge computing is defined as distributed computing paradigm that has network, compute, and storage capacity at the edge of the network through small data centres closer to the users. The current landscape integrates Internet of Things (IoT) devices with the cloud through the edge computing layer by filtering, preprocessing, and aggregating data generated by IoT devices [6]. The edge computing layer includes edge nodes, actuators, sensors geographically distributed on the edge layer. They provide mobility, location-awareness, interoperability, heterogeneity, and real-time responses to applications deployed over the edge computing layer. This potentially brings several advantages like lower latency, less bandwidth utilization, and increased energy efficiency. Overall, this makes edge computing appropriate for computation-intensive and latency-sensitive applications. There are also few drawbacks that are acknowledged in the existing literature like low individual computing power, unreliable devices and limited load balancing capacity[7, 8]. Hence, the pros and cons must be considered while developing edge-based applications. In Addition, the access technology in edge computing is mainly through wireless gateways with 1-hop distance Wi-Fi or cellular connections. In various scenarios, autonomous communication such as Machine-to-Machine (M2M), Device-to-Device (D2D), Car-to-Infrastructure (C2I), Car-to-Car (C2C) is used among users and devices continuously [9]. The communication scenarios and the distributed features of edge computing make it vulnerable to many security threats and issues.

Edge computing is subjected to several security challenges. It can be prone to many malicious activities like malicious access or manipulation of sensitive data. Like, the wireless networking technology can be broken by malicious users without physical contact to determine the resource location and access the data. Finally, limited resources make it challenging to adopt enhanced data protection models [10]. The above challenges motivate a need for developing an efficient data security system in the edge computing layer. Although most threats are inherited from the cloud, breaking network access, unknown stakeholders or determining resource locations are specific threats that arise in edge computing. This demands a particular study on data security issues in edge computing. It can be noted that [11–13] are complete reviews on data security, privacy, and trust issue, and there are no reviews, particularly on data integrity issues. In edge computing, computation details are not transparent like a cloud; malicious users can unknowingly tamper, hide, or remove data. More importantly, users are unaware of the stakeholders processing it. This affects data integrity and disturbs the edge data analysis. Thereby this review focuses mainly on data integrity issues and identifies the shortcomings in existing work.

In the remainder of this paper, an overview of edge data analytics and security issues arising from data analysis are discussed. Later a review on data integrity issues is analyzed, and research gaps are identified. Finally, research directions towards developing a secure data framework are proposed.

### 1.1 Edge Data Analytics

Edge data analytics (EDA) includes a data stream (DS) from the ubiquitous devices sent to edge nodes (EN) that are typically close to their sources. The communication channel at the proximity ensures that data is processed without delay. Each edge node maintains appropriate data processing models to update the data analysis and infer real-time interactions to the applications. The data models transmit processed data to the nearby edge server. This approach includes the parallel processing of data streams to foster the scalability of the system. The processed data are aggregated at the edge servers (ES) and broadcasted to all the edge nodes for accurate inference. The generic data analysis process is illustrated in Fig. 1 [14]. The security threats during this process can affect data quality and accuracy. While data is processed on the edge nodes, data owners lose complete control over their data. This can also lead to malicious intruders gaining control over the network. Finally, few infringements here can cause data theft or tamper with storage devices [11]. Therefore, while benefiting from the edge computing paradigm from various IoT applications, ensuring security to data present at the edge layer becomes the most valuable research in secure edge data analytics.

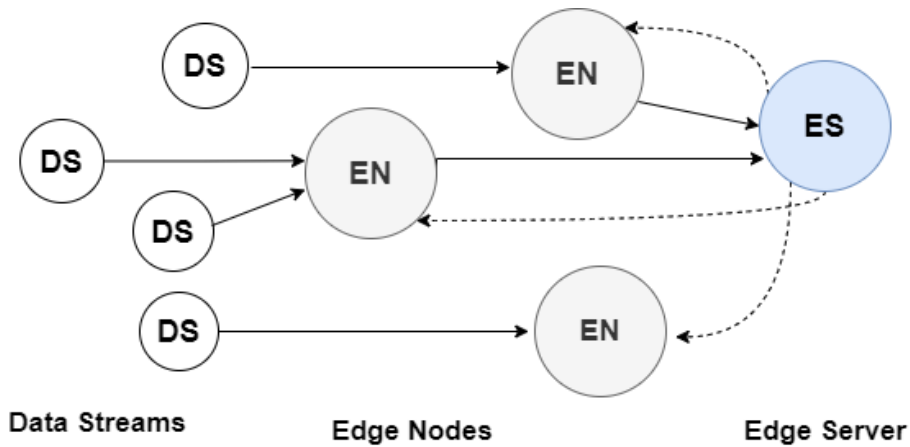

**Fig. 1.** Edge Data Analytics

Observation #1: During data traversal from IoT devices to edge servers, end-users lose complete ownership of their data. In case of a security breach, their data can be mishandled by intruders. Lack of service agreements like in cloud computing can create uncertainty among end-users in this regard.

## 2 Data Security issues

Securing data from external threats is an evolving research area since the early days of computers and their usage. The primary concern while securing data is to ensure reliability, consistency, and accessibility to the authorized personnel. To achieve this, the CIA triad is widely used in information technology. The origin of the CIA triad is deeply rooted in the military security system that focuses on protecting internal information from external threats. In academia and information security research, this approach is an important asset for protecting information [15]. The three categories of the CIA triad that are the foundation for the data protection system are: confidentiality, integrity, and accessibility and is represented as shown in Fig. 2. They are identified as follows:

1. **Confidentiality** – It is the level of privacy that is required at each level of data processing. The attacker can observe the data pattern through traffic analysis and infer the data. Monitoring unauthorized supervision of end-users data is a security measure to achieve confidentiality.
2. **Integrity** – It is a kind of data sabotage where attackers alter or modify intentionally. Compromising data integrity can be achieved mainly during data outsourcing, where data owners lose complete control. Data validity against undesired changes assures integrity.
3. **Availability** – It is a way of preventing authorized users from referring to or modify data. The attacker can hack the data resources and gain complete control over them. Data must be reachable to required processing nodes to ensure availability.

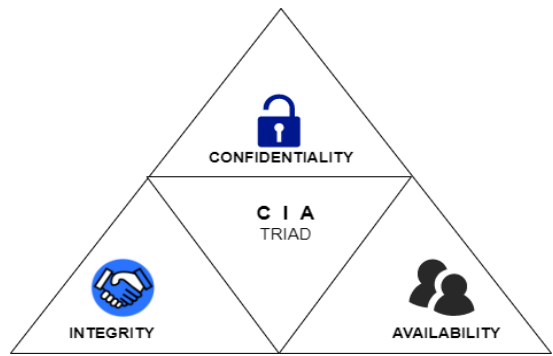

**Fig. 2.** Data Security Triad – CIA

Many critical security threats are growing concerns in information technology. But attacks that affect data integrity can have a severe effect that can mislead the data analysis process and deceive real-time interaction provided by the edge nodes. Therefore, addressing data integrity attack is challenging and crucial to establish trust between the end-users and the application running on edge computing. A complete understanding of the system with intelligence is required to make this attack effective [16].

**2.1 Data integrity issues**

The goal of the data integrity attack is to mislead the operator with erroneous data. To achieve these, attackers can create malicious data through spam or false data injection. Hence, in this section, existing security mechanisms that address these threats are analyzed to identify their pros and cons and tabulated in Table 1. Spam can be defined as sending an irrelevant, inappropriate and unsolicited message to several users to exploit sensitive information [16]. Fog augment, and Machine Learning (ML) based spam detection tool detects spam in both incoming and outgoing Short Message Service (SMS). It is an efficient tool to identify spam SMS that is illegitimately installed on mobile devices due to malware. Various ML classifiers and data preprocessing termed as a filter in this tool assist significantly in identifying spam on mobile devices and cloud servers. The fog computing layer is used only to perform classification and identification and has no specific role in data processing [17]. Deep learning-based multistage and elastic spam detection on mobile social networks like Twitter, Facebook, and Sina Weibo, significantly improve real-time quality of service. Initial detection is performed on mobile terminals, and later the result with the message is passed to the cloud server. The server extensively computes to identify and confirm the spam through the detection queue. The computational time and resources in the detection queue may incur superfluous resources [18].

Wi-Fi enabled a mesh of cloudlets to secure data during transmission, storage and processing on mobile devices and remote clouds. This framework includes hierarchical intrusion detection and prevents spam/virus attacks between cloud, edge, and mobile devices. Collaborative intrusion detection and MapReduce spam filtering process does predictive security analyses at the backend, emphasizing real-time response towards intrusions through the trusted remote cloud [19].

Data injection is an attack akin to spam where legitimate data is modified, altered, or injected with false data to affect the data processing efficiency. Paillier homomorphic encryption scheme is used to secure data from false injection and modification attacks on the edge computing layer. This includes three phases in terms of key generation, encryption, and decryption. Without having a third-party trust assigning authority, data privacy is obtained in the edge computing layer. This approach is more efficient than traditional methods, but when the number of IoT devices increases, the edge nodes' computation overhead increases gradually [20]. S-CLASSIFIER is a false data injection detector on edge nodes. It includes four components – data controller to transmit sensory data, indicator to indicate connectivity between nodes and sensors, aggregator to receive collective data from the data collector and finally, a detector responsible for classifying the data and deciding whether it is good or bad. In case of detecting bad data, the detector issues an alarm to the other node, blocking them from further data transmission. The Support Vector Machine (SVM) method makes this method simple and efficiently divides train data on edge nodes [21]. Comparatively, a third-party auditor is used to audit data integrity through a homomorphic authenticator. Bilinear pairing keys are used along with the homomorphic authenticator for auditing multimedia data. This includes seven phases of key generation, data uploading, backup, recovery, and auditing. This is later outsourced for a third-party auditor to validate storage proof using a public key. Although this approach is efficient in recovering data, resource consumption and time computation may be more for this redundant process [22].

There are many other methods that are suitable for smaller attack areas [23], specific models for healthcare applications [24] [25] or a model with only conceptual design [26]. These recent advances show that addressing data integrity attacks is a hot topic in edge computing. Since these attacks can obstruct the normal operation of the network without being detected, the intention of these attacks can be to gain control over the system, establish undetectable errors, or malicious intent of exploiting the system. Therefore, addressing these attacks is crucial to ensure data privacy and confidentiality. Additionally, most existing methods include complex computations, expensive pre-computation steps, multiple execution of single steps, or outsourcing computation to third-party authorities. These approaches for resource and computation limited edge computing can be challenging and affect the data processing efficiency. It can also be observed that identifying any form of intrusion when the data enter the edge layer can greatly reduce the intensity of the attack. Overall, there is a need for secure data framework that can detect the intrusion at an early stage and diminish the attack without adding complex computational load on the edge nodes.

**Table 1.** Review on existing security mechanism to address integrity issues in Edge Computing

| Ref | Method | Evaluation | Pros/Cons |
|---|---|---|---|
| [17] | Fog Augmented and ML-based SMS spam detection system generated by malware on mobile devices | 5 ML classifiers and filters are used to identify spam on mobile devices and cloud servers. Based on the user preference, appropriate spam filters are recommended | The data preprocessing performed is termed as a filter, which assists greatly in identifying spam. |
| [18] | Deep learning-based multistage and elastic spam detection in mobile social network | A server's detection queue with a timer identifies and elastically abandon the spam data received from mobile terminals. Edge computing handles computational resources for multistage detection | Real-time message detection significantly improves the service, but multistage detection consumes enormous resources |
| [19] | Wi-Fi enabled mesh of cloudlets to secure cloud by pervasive mobile users | Hierarchically trusted secure framework emphasizing to remove malicious attacks and filter spam from mobile devices | Collectively performs intrusion detection and spam reduction. Effectively offload task between cloud and cloudlets |
| [20] | Paillier homomorphic encryption scheme | Protects sensitive information through 3 stage cryptographic | Reliable and efficient security framework to ensure data integrity in edge |

| | against data injection attack | operations and uses blinding factors to enhance data privacy | computing. However, it increases the computational load on edge devices |
|---|---|---|---|
| [21] | S-CLASSIFIER an edge node detector to identify false data injection in smart grid | ML-based classifier identifies the attack during data transmission, and sub-controller on the edge nodes takes the decision immediately to mitigate the attack by issuing an alarm | Edge nodes detect the attack at a faster rate and prevent it |
| [22] | Homomorphic authenticator to audit data integrity on multimedia data | Stores historical data in a one-way lined information table. Third-party auditor audits the stored data to ensure data integrity, recovery and recure from replay attack | Significantly support data recovery and backup in the cloud and determines it can be applied in a real-time scenario for multimedia security |
| [23] | Constraint-based false data injection model | Linear and nonlinear based two models are used to identify a complicated and straightforward attack. | Better for smaller attack area |
| [24] | Data integrity preservation in healthcare applications | A transaction dependency graph is used to monitor the transaction and identify intrusions initially. Later, using a damage assessment algorithm on serializable history, data integrity is verified | Intrusion detection significantly monitors unauthorized access in the edge layer |

Observation #2 – Complex security frameworks provides security to the edge framework but overload the edge nodes. Thereby it makes it challenging for the edge nodes to retain their characteristics, especially the real-time responsiveness. This can significantly affect the future adoption of edge computing in any application.

Observation #3 – Cryptographic techniques, outsourcing validation externally, and maintaining multiple logs increase the edge nodes' time and resource utilization through frequent interactions. These interactions unknowingly can create an opportunity for the intruders to let them understand the system's working.

## 3      Future Research Directions

Future research directions that could leverage the existing solutions to achieve data integrity in the edge layer are listed below:

- **Designing reputation management model of edge nodes** – The decentralized edge computing framework collect, process, and analyze data frequently on the edge layer. As noted in observation #1, edge data analytics lack users' confidence in their data. The potential loopholes are mainly due to malicious users gaining access to the network. Therefore, creating fairness among the participating nodes and users can patch these loopholes [27]. A scalable reputation-based security mechanism can be more suitable considering historical interaction and frequent peers' updates.
- **Adopting Federated Learning algorithms** – Edge computing is the most feasible technology due to its local computation to reduce data intensity on the cloud. As mentioned in observation #2, overloading edge nodes can affect the feasibility gradually. Federated Learning, a distributed machine learning approach, facilitates training large amounts of data on ubiquitous devices. This approach significantly improves the data analysis process; thereby, adopting this can complement edge computing and address the fundamental problem of privacy, ownership, and locality data.
- **Adopting Fine-Grained data consistency enhancing model** – Following observation #3, the multiple interactions due to complex cryptography or third-party security measures leads intruders to gain access to the network. Adopting a straightforward fine-grained data consistency model can reduce the interactions between the nodes and achieve secure collaboration. This approach is proved to be efficient in cloud computing [28]. Therefore, adopting a similar approach can provide data tracking and enhance interoperability between heterogenous edge nodes.

## 4      Conclusions

With the proliferation of IoT devices, edge computing is becoming an emerging technology to manage ample data through local edge data analytics and offering resources. Alongside the security concern due to proximity, data handling is also progressing. Therefore, there is a paramount need to address the data security issue to sustain edge computing. This paper addresses security issues that affect data integrity, influencing data efficiency on the edge layer. Existing security methods available to address this are reviewed to identify the gaps in current research. Lastly, future research directions towards data security, privacy and secure data analytics are directed.